\documentclass[conference]{IEEEtran}
\IEEEoverridecommandlockouts
\usepackage{cite}
\usepackage{amsmath,amssymb,amsfonts}
\usepackage[normalem]{ulem}
\usepackage{algorithmic}
\usepackage{graphicx}
\usepackage{textcomp}
\usepackage{tgcursor}  
\usepackage{epigraph}
\usepackage{tikz}
\usepackage{xcolor}
\usepackage[mathcal]{eucal}
\usepackage[shortlabels]{enumitem}
\usepackage{tablefootnote}
\usepackage{lipsum,afterpage,refcount}
\usepackage{hyperref}
\def\BibTeX{{\rm B\kern-.05em{\sc i\kern-.025em b}\kern-.08em
    T\kern-.1667em\lower.7ex\hbox{E}\kern-.125emX}}
\begin{document}
\newcommand{\setfootnotemark}{%
  \refstepcounter{footnote}%
  \footnotemark[\value{footnote}]}

\title{TINKER: A framework for Open source Cyberthreat Intelligence
%\thanks{Supported by IBM AI Research Collaboration.}
}

\author{\IEEEauthorblockN{1\textsuperscript{st} Nidhi Rastogi}
\IEEEauthorblockA{\textit{Department of Software Engineering} \\
\textit{Rochester Institute of Technology}\\
Rochester, NY, USA \\
nidhi.rastogi@rit.edu}
\and
\IEEEauthorblockN{2\textsuperscript{nd} Sharmishtha Dutta}
\IEEEauthorblockA{\textit{Department of Computer Science} \\
\textit{Rensselaer Polytechnic Institute}\\
Troy, NY, USA \\
duttas@rpi.edu}
\and
\IEEEauthorblockN{3\textsuperscript{rd} Alex Gittens}
\IEEEauthorblockA{\textit{Department of Computer Science} \\
\textit{Rensselaer Polytechnic Institute}\\
Troy, NY, USA \\
gittea@rpi.edu}
\and
\IEEEauthorblockN{\hspace{40mm}4\textsuperscript{th} Mohammed J. Zaki}
\IEEEauthorblockA{\hspace{40mm}\textit{Department of Computer Science} \\
\textit{\hspace{40mm}Rensselaer Polytechnic Institute}\\
\hspace{40mm}Troy, NY, USA \\
\hspace{40mm}zaki@cs.rpi.edu}
\and
\IEEEauthorblockN{5\textsuperscript{th} Charu Aggarwal}
\IEEEauthorblockA{\textit{}
\textit{IBM Research}\\ 
T. J. Watson Research Center\\
Yorktown Heights, NY, USA \\
charu@us.ibm.com}
}

\maketitle

\begin{abstract}
Threat intelligence on malware attacks and campaigns is increasingly being shared with other security experts for a cost or for free. Other security analysts use this intelligence to inform them of indicators of compromise, attack techniques, and preventative actions. Security analysts prepare threat analysis reports after investigating an attack, an emerging cyber threat, or a recently discovered vulnerability. Collectively known as cyber threat intelligence (CTI), the reports are typically in an unstructured format and, therefore, challenging to integrate seamlessly into existing intrusion detection systems. This paper proposes a framework that uses the aggregated CTI for analysis and defense at scale. The information is extracted and stored in a structured format using knowledge graphs such that the semantics of the threat intelligence can be preserved and shared at scale with other security analysts. Specifically, we propose the first semi-supervised open-source knowledge graph-based framework, TINKER, to capture cyber threat information and its context. Following TINKER, we generate a Cyberthreat Intelligence Knowledge Graph (CTI-KG) and demonstrate the usage using different use cases.
\end{abstract}


\section{Introduction}
Security analysts investigate cyber threats to organizations and regularly determine remedial actions to prevent them in the future. They gather cyber threat intelligence analysis reports that describe newly emerging threats and/or ways to mitigate them. This information, called Cyber Threat Intelligence (CTI), offers a detailed summary of one or more cyber attacks and aggregates relevant knowledge and context. It covers information on zero-day attacks, newly identified vulnerabilities, indicators of compromise (IoC), threats, and attack patterns. Such information is multi-modal, predominantly unstructured, and does not follow a specific format. Other cyber threat analysts utilize shared threat intelligence (free or paid) to mitigate potential threats and defend their organizations. Existing unstructured and structured sources of threat information, such as analysis reports from security organizations, blogs, common vulnerabilities, and exposure databases (CVE), provide curated, sometimes very detailed, accounts of the current cyber threat landscape. Security companies such as McAfee and Symantec, researchers, government agencies, and experts publish these accounts largely in unstructured text. MITRE, NIST, OASIS Open have spearheaded initiatives that make security information available in structured formats through standards like CVE/NVD and STIX.

% \begin{figure}[htbp]
% \centering
% \frame{\includegraphics[width=.43\textwidth, height=0.18\textwidth]{figures/threatReportSnippet.png}}
% \caption{Excerpt of a CTI report for malware family \textit{Backdoor.Winnti}\cite{broadcomReport}.}
% \label{fig:report_snapshot}
% \end{figure}

\textbf{T}hreat \textbf{In}telligence \textbf{K}nowledge \textbf{E}xtracto\textbf{r} (TINKER) addresses these two main gaps: (a) whereas the current taxonomy and standards provide extensive coverage of threat concepts, they lack semantic and contextual associations and (b) threat intelligence extraction models not trained on cyber threat corpus can lead to capturing fragmented and sometimes fallacious threat information.
\par
TINKER transforms multi-modal CTI into a structured format while preserving the context of the information. 
% TODO LATER: For example, \textcolor{red}{text from threat report that is used in Fig1, annotated, and can be confused}. 
TINKER uses ontologies and information extraction models to capture CTI. It seamlessly integrates heterogeneous sources of data, captures data and its context, and structures them using the \textbf{C}yber\textbf{t}hreat \textbf{I}ntelligence \textbf{K}nowledge \textbf{G}raphs (CTI-KG), which enables further machine learning like analysis~\cite{unknown-author-2022}. This paper focuses on extracting threat intelligence from the generic security corpus. In this paper, we make the following contributions:
\begin{enumerate}
\item We present TINKER, the first end-to-end framework for capturing and analyzing multi-modal threat intelligence information using a semi-supervised approach.
\item Using TINKER, we generate an open-source CTI-KG, the first open-access knowledge graphs for general cyber threat intelligence.
\item We demonstrate the use of CTI-KG through two use cases to infer previously unknown threat information such as vulnerabilities, associations between malware, and malware author.
\item We present an open-access corpus of over 52K triples comprising 30k unique named entities and 22 relationships. This is a new benchmark framework for contextual cyber security research problems.
\end{enumerate}

\section{Definitions}
\textbf{Cyberthreat Intelligence Knowledge Graph (CTI-KG)} is a directed multi-modal graph. Related definitions are given in the Table~\ref{table:notations}. All triples have three vectors,  $\mathbf{e_1, e_2} \in \mathbb{R}^{d_e}$, $r \in \mathbb{R}^{d_r}$.

\begin{table}[!h]
\scriptsize
\renewcommand{\arraystretch}{1}
\caption{Notations and Descriptions}
\label{table:notations}
\centering
\begin{tabular}{|p{1.9cm}|p{6cm}|}
\hline
Notation & Description\\
\hline
$\mathcal{E},{R},{T}$ & set of all entities, relationships and triples respectively\\
$\langle e_{head},r,e_{tail}\rangle$ & head entity, relationship and tail entity\\
$\langle \textbf{h, r, t}\rangle$ & embedding of \textbf{h}, \textbf{r} and \textbf{t}\\
$d_e$, $d_r$  & embedding dimension of entities, relationships\\
$n_e$, $n_r$, $n_t$ & number of \textbf{h+t}, \textbf{r}, triples\\
$f_\phi$ & scoring function of each triple $\langle \textbf{h, r, t}\rangle$. measures plausibility of a fact in $\mathcal{T}$, based on translational distance or semantic similarity \cite{li2020survey}\\
\hline
\end{tabular}
\end{table}

\textbf{Ontological mapping} is performed using classes and relationships defined in cyberthreat ontologies~\cite{rastogi2020malont}\cite{swimmer} to extract triples from unstructured text. An ontology maps disparate data from multiple sources into a shared structure that maintains a logic in that structure using rules and follows a common vocabulary~\cite{BuildingAnOntologyOfCS}. For example, there are classes named Malware and Location in MALOnt. According to the relationships defined in an ontology, an entity of \textsf{Malware} class can have a relationship ``similarTo'' with another \textsf{Malware} class entity. However, according to the rule engine, an entity belonging to the class \textsf{Malware} can never be related to an entity belonging to the class \textsf{Location}. The same two entities can have multiple relationships between them. For example, an entity of type \textsf{Malware} and \textsf{Application} can have relationships such as~\textsf{uses, communicates\_with} have very different meaning. Therefore, context plays a significant role in analyzing entities.

\section{TINKER Framework} \label{tinker}
\begin{figure}[!h]
    \centering
   \frame{\includegraphics[width=.45\textwidth]{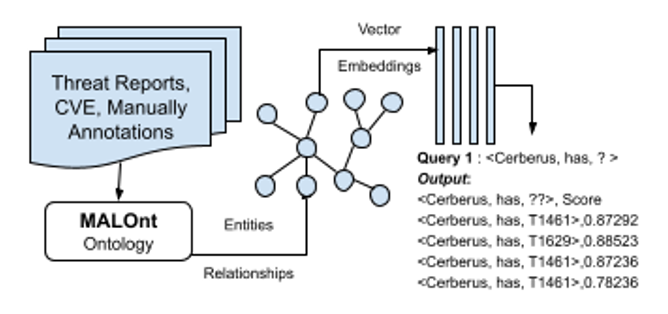}}
    \caption{System Architecture for CTI-KG construction and showing entity prediction.}
    \label{fig:SystemArchitecture.png}
\end{figure}

We describe TINKER, the framework for generating CTI-KG and inferring threat intelligence (see Fig.\ref{fig:SystemArchitecture.png}). The framework comprises modules that perform semi-supervised data collection using information extraction models, CTI-KG construction, generating vector embeddings from triples, and inference. 
%\subsection{Framework Modules}\label{modules}
\textbf{Automated data collection}: We aggregate hundreds of reports on threat intelligence uploaded on GitHub repositories. These reports are written between the years 2010-2022 by security analysts from top security companies such as FireEye, Symantec, McAfee, Microsoft Research, security bulletins, TechCrunch, and ESET. Not all reports are consistent, accurate, or comprehensive and can vary in technical depth and breadth of the attack description.
\newline
\textbf{Information Extraction}: This module is responsible for extracting contextual and structured cyber threat information. Named entity extraction (NER) and the semantic relationship between them (relation extraction, RE) from unstructured text form triples -- the fundamental unit of CTI-KG. We first create ground truth using a combination of hand-annotated CTI, supervised and semi-supervised NER, and relationship extraction. For instance, consider the following snippet from a CTI report\cite{theLastWatchdog}.

\textit{\footnotesize{``... DUSTMAN can be considered as a new variant of ZeroCleare malware ... both
DUSTMAN and ZeroCleare utilized a skeleton of the modified “Turla Driver Loader (TDL)” ... The malware executable file ``dustman.exe” is not the actual wiper. However, it contains all
the needed resources and drops three other files [assistant.sys, elrawdisk.sys, agent.exe] upon execution...”}}. 
\begin{table}[!h]
%% increase table row spacing, adjust to taste
\renewcommand{\arraystretch}{1.3}
\caption{Triples corresponding to the CTI report.}
\label{table:triple_list}
\scriptsize
\centering
%% Some packages, such as MDW tools, offer better commands for making tables
%% than the plain LaTeX2e tabular which is used here.
\begin{tabular}{|c| c| c|}
\hline
Head Entity & Relationship & Tail Entity\\
\hline
$\langle$DUSTMAN, & similarTo, & ZeroCleare$\rangle$\\
$\langle$DUSTMAN, & involves, & Turla Driver Loader(TDL)$\rangle$\\
$\langle$ZeroCleare, & involves, & Turla Driver Loader(TDL)$\rangle$\\
$\langle$DUSTMAN, & involves, & dustman.exe$\rangle$\\
$\langle$dustman.exe, & drops, & assistant.sys$\rangle$\\
$\langle$dustman.exe, & drops, & elrawdisk.sys$\rangle$\\
$\langle$dustman.exe, & drops, & agent.exe$\rangle$\\
\hline
\end{tabular}
\end{table}
%\vspace{-4.00mm}

\noindent
\textbf{CTI-KG Construction}: Together, thousands of triples construct CTI-KG where the entities and relationships model nodes and directed edges, respectively. A knowledge graph contains 1-to-1, 1-to-n, n-to-1, and n-to-n relationships between entities.
\newline
\textbf{Inferring Threat Intelligence}: This task infers threat intelligence about an emerging or existing security threat along with a confidence score (0.0-1.0). We discuss various embedding models we explored for the inference task in the related work section \ref{relatedWork}. For this paper, we adapt the TuckER~\cite{balavzevic2019tucker} model, a linear model based on  Tucker decomposition of the binary tensor representation of CTI-KG triples. TuckER creates vector embeddings for all triples in CTI-KG and infers missing entities for classes defined in the ontology~\cite{rastogi2020malont}.
\par
\textbf{Evaluation Criteria: }Knowledge Graph-based threat intelligence is a new paradigm for inferring information about emerging threats. Unlike machine learning-based models that predict a value or label a data point based on prior learning from training data, knowledge graph-based prediction provides a list of entities inferred by a triple, using the vector embedding approach and the confidence score. Fundamentally, inferring threat intelligence entities is modeled in CTI-KG by predicting the missing entities in the graph. The training step of the CTI-KG constructs the KG. Further, in the inference step, we perform the inference operations to conclude new facts not explicitly present in CTI-KG.
\par
For example, consider ``Adobe Flash Player'' as an entity belonging to the class \textsf{Software} in CTI-KG. Several entities have an incomplete triple with a missing tail entity in the training phase because that information was missing from the threat reports. This triple will be of the form $\langle$\textit{Adobe Flash Player, hasVulnerability, \textbf{?}}$\rangle$. The entity inference operation will infer a list of missing tail entities and the confidence score of the predictions, i.e., a vulnerability of the Software titled ``Adobe Flash Player.'' In another example, for an incomplete triple, where the tail entity is the malware family ``Hydraq'', the CTI-KG returns the class of the head entity, \textsf{Campaign}, and the entity itself, \textsf{Trojan.Hydraq} for the query $\langle$\textit{\textbf{?}, involvesMalware, Hydraq}$\rangle$. In CTI-KG, entity inference is evaluated, and confidence score is computed as described below~\cite{wang2017knowledge}:

\begin{enumerate}

\item When is an incomplete triple, $\langle$\textit{\textbf{h, r, t}}$\rangle$ needs to be completed by an entity, a set of candidate entities are generated along with the scoring function, $f_\phi(h,r,t)$, for each triple. The score is sorted to obtain the rank of the true test triple in the ordered set. This rank evaluates the performance of the model in predicting $tail$ given  $\langle$ \textit{head, relation, ?} $\rangle$. A higher rank indicates a model's better efficiency in learning embeddings for the entities and relations.
\item \textit{Hits@n} is calculated from the ranks of all true test triples, where \textit{Hits@n} denotes the ratio of true test triples in the top \textit{n} positions of the ranking. Smaller values of \textit{Hits@n} indicate a better model when comparing embedding models for prediction.
\end{enumerate}

\section{Dataset Preparation}\label{dataset}
The first dataset comprises hand-annotated triples, and the second is from a larger CTI corpus from various cyber threats such as APT, malware, and phishing. CTI report is originally in .pdf or .html format and then converted into text format. Pre-processing includes several techniques to minimize noise and remove unnecessary information from the text. We tokenize and parse the plain text sentences into words with start and end positions using a python library called spaCy, which allows POS tagging, dependency parsing, and word vectors. Each token is assigned an id and stored for each CTI report.
\newline
\textbf{Ground Truth}: 85 reports were hand-annotated by mapping entities and relationships to classes defined by ontology. Two graduate and five undergraduate student researchers with cybersecurity education performed this task in 3 to 4 months. A senior, more experienced full-time researcher oversaw the task and broke ties when needed. Later, the same expert verified the hand-annotated triples. The CTI reports were written between 2010-2022. The comparison between the structure of the knowledge graphs generated from the hand-annotated ground truth and CS40K dataset, their properties, and corpus details are described in Table \ref{table:dataset}. The columns in the table compare the number of entities, relations, and triples, the average degree of entities, and graph density of our datasets (CS3K, CS40K) and benchmark datasets.The average degree across all relationships and graph density are defined as $n_t/n_e$ and  $n_t/n_e^2$, respectively \cite{padia2019knowledge}. A lower score for these two metrics indicates a sparser graph.
\par
\textbf{Larger Dataset}: A total of 1,100 threat advisory reports were downloaded from GitHub repositories. They were written by analysts from major security organizations' websites such as Symantec, Kaspersky, Microsoft, IBM, FireEye, TrendMicro, and McAfee. The annotated key phrases were classified into entities derived from semantic categories defined in cyber threat ontologies \cite{rastogi2020malont},\cite{swimmer}. The inference rule engine from the ontologies was applied when forming relationships between entities to produce triples used to build the knowledge graphs. We use semi-supervised learning to train NER and relationship extraction models by providing a few hand-annotated samples. The CTI-KG comprises 40,000 triples generated from 27,354 unique entities and 11 relations.
\par

\begin{table}[!t]
\renewcommand{\arraystretch}{1.2}
\caption{Description of Datasets}
\label{table:dataset}
\setlength{\tabcolsep}{4.8pt}
\small
\centering
\begin{tabular}{|p{1cm}|p{.8cm}|p{.3cm}|p{1cm}|p{.8cm}|p{.9cm}|p{1.2cm}|}
\hline
Dataset & $n_e$ & $n_r$ & $n_t$ & docs & avgDeg & density\\
\hline
\textbf{CS3K} & 5,741 & 22 & 3,027 & 81 &0.5273&0.00009\\
\hline
\textbf{CS40K} & 27,354 & 11 & 40,000 & 1,100&1.46& $5.34 \times 10^{-5}$\\
\hline
\end{tabular}
\end{table}

\textbf{Information Extraction and Ontological Mapping: }Concepts that a generic named entity recognition model can easily classify require training using a representative dataset. We train several NER models with suggestive keywords and the context within which the keyword occurs. We train an odd number of NER models for each class. Since the entity extraction models are running independently, it was likely for close class types to have overlapped. We resolved this issue by following these rules, (a) we compared the confidence score of entities occurring at the same position in the same document and between classes, (b)for the same confidence scores, we marked those classes as ambiguous. We retained these instances until the relationship extraction phase. At that time, we relied on the rule engine from the ontology to determine the correct class of the head and tail entity for a given relationship (described in the next section). The entity classification (EC) step addresses the disambiguation of an entity classified into more than one semantic category and assigns only one class based on an empirically learned algorithm. EE and EC maximize the automation of an otherwise exhaustive entity recognition process. Semantic text patterns from NER and RE map to classes (called entities in KG) and object properties (called relationships in KG) defined by an ontology. For example, in~\cite{rastogi2020malont,christian2021ontology}, three classes largely describe malware behavior -- \textsf{Malware}, \textsf{Vulnerability}, and \textsf{Indicators of Compromise}. We segregate tokenized words' extraction into domain-specific key-phrase extraction and generic key-phrase extraction (\textsf{class:Location}). The cybersecurity domain-specific EE task was further subdivided into factual key-phrase extraction (\textsf{class:Malware}) and contextual key-phrase extraction (\textsf{class:Vulnerability}). We use precision and recall measures as evaluation criteria to narrow down the EE models for different classes. For example, the Flair library~\cite{akbik2019flair} gave high precision scores (0.88-0.98) for the classification of generic and factual domain-specific key phrases such as - \textsf{Person, Date, Geo-Political Entity, Product (Software, Hardware)}. The diversity in writing style for the CTI reports and the noise introduced during reports' conversion to text led to different precision scores for classes but within the range of 0.9-0.99. For contextual entity extraction, we expanded the SetExpan \cite{shen2017setexpan} model to generate all entities in the text corpus. The feedback loop that inputs seed values of entities based on confidence scores allowed for improving the training dataset. Some semantic classes used in this approach are \textsf{Malware, Malware Campaign, Attacker Organization, Vulnerability}. We perform relationship extraction for CTI-KG using a distantly supervised artificial neural network model \cite{yao2019docred} pre-trained on the Wiki data and Wikipedia, initially with three features. The CS3K triples and text corpus form the training dataset and the entities generated in the previous steps, and the processed text corpus is the testing dataset. We built on this learning and improved automatic relation extraction (RE) by using semi-supervised relation extraction methods\cite{hu2021semi} aimed to leverage unlabeled data in addition to learning from limited samples.

% \begin{figure}[htp]
% \centering
% \frame{\includegraphics[width=.49\textwidth, height=0.18\textwidth]{figures/brat.jpg}}
% \caption{Annotation using Brat annotation tool.}
% \label{fig:annotate}
% \end{figure}

% \begin{figure}[hbt!]
%         \centering
%         \frame{\includegraphics[width=0.22\textwidth, height=0.24\textwidth]{figures/classes.png}}
%         %\hspace{.5cm}
%         \frame{\includegraphics[width=0.23\textwidth, height=0.24\textwidth]{figures/properties.png}}
%         \hspace{.5cm}
%         \caption{Main Classes (left), Object Properties (right) from a cyber threat ontology. Instances of classes, called entities and instances of object properties called relationships are mapped to CTI text using NER and relation extraction models}
%         \label{fig:ontology}
% \end{figure}

\section{Experiment and Evaluation}\label{experiment}

We infer threat intelligence on the CTI-KG through a tensor factorization-based approach~\cite{balavzevic2019tucker} using PyTorch. To train all the datasets, we choose all hyper-parameters based on validation set performance through random search. Since all the datasets contain a significantly smaller number of relationships relative to the number of entities, we set entity and relationship embedding dimensionality to $d_e= 200$ and $d_r=30$, learning rate = 0.0005, batch size of 128 and 500 iterations. We use batch normalization and dropout to decrease training time. For evaluation, for each test triple, we generate $n_e$ candidate triples by combining the test entity-relation pair with all possible entities $\mathcal{E}$, followed by ranking the computed scores. We use evaluation metrics standards across the link prediction literature, and they are described in Section \ref{tinker}.
\par
CS3K validation set is used for tuning the hyper-parameters for CS40K. For the prediction experiment on CS3K, we achieved the best performance with $d_e=200$ and $d_r=30$, learning rate = 0.0005, batch size of 128 and 500 iterations. We split each dataset into  70\% train, 15\% validation, and 15\% test data. Due to the limited hand-annotated dataset, we tested the CS40K dataset using CS3K. The challenge with a smaller ground truth can lead to the overfitting of results. To increase the size of the ground truth, we increase the size of triples through a semi-supervised method, i.e., by the process described earlier. A human-in-the-loop (a security expert) is employed to fix the false positives manually.

\subsection{Inferring Threat Intelligence from CTI-KG}\label{overallResults}

Table \ref{table:result-benchmark3} displays the results of entity prediction existing embedding models on the benchmark datasets. An upward arrow ($\uparrow$) next to an evaluation metric indicates that a larger value is preferred for that metric and vice versa. For ease of interpretation, we present \textit{Hits@n} measures as a percentage value instead of a ratio between 0 and 1. TransE has the largest \textit{Hits@10}, whereas TransR performs better on the rest of the three datasets. However, TransH is more expressive (capable of modeling more relations) than TransE despite having the simplicity of TransE (time complexity $\mathcal{O}(d_e)$). The space complexity of TransR ($\mathcal{O}(n_ed_e+n_rd_ed_r)$) is more expensive than TransH ($\mathcal{O}(n_ed_e+n_rd_e)$)\cite{wang2017knowledge}. Additionally, it takes six times more time to train TransR(on benchmark datasets) than TransH\cite{lin2015learning}. Considering these model capabilities and complexity trade-offs, we choose TransH as a representational translation-based model. DistMult performs better than TuckER by a small margin of 0.37\% in only one case (\textit{Hits@10} on WN18). TuckER and ComplEx both achieve very close results on multiple cases. 
\par
Therefore, we look closely at the results of the recently refined datasets (FB15K-237 and WN18RR). The CS3K dataset shares similar properties with these two datasets (CS3K has hierarchical relationships as in WN18RR and no redundant triples as in FB15K-237). Since the space complexity is very close (ComplEx: $\mathcal{O}(n_ed_e+n_rd_e)$ and TuckER: $\mathcal{O}(n_ed_e+n_rd_r)$)  we choose TuckER for evaluation on CS3K, as it subsumes ComplEx \cite{balavzevic2019tucker}.

We evaluate the performance of TransH~\cite{wang2014knowledge} and TuckER \cite{balavzevic2019tucker} on our datasets. Tables \ref{table:result:CS3K} and \ref{table:result:CS40K} display the performance of TransH and TuckER on CS3K and CS40K. TransH's MRR score on CS3K improved when compared to the benchmarks. Tucker's performance improved in all measures when compared to the other benchmark results. TuckER outperforms TransH by a large margin on the CS3K dataset. 

\par Since TuckER performed significantly better than TransH on the ground truth dataset, we employ only TuckER on the CS40K dataset. It is noteworthy that all the measures except mean rank (MR) improved on CS40K compared to CS3K. It confirms the general intuition that the model learns the embeddings more effectively as we provide more data. MR is not a stable measure and fluctuates even when a single entity is ranked poorly. We see that the MRR score improved even though MR worsened, which confirms the assumption that only a few poor predictions drastically affected the MR on CS40K.
\textit{Hits@10} being 80.4 indicates that when the entity prediction model predicted an ordered set of entities, the true entity appeared in its top 10 80.4\% times. \textit{Hits@1} measure denotes that 73.9\% times, the model ranked the true entity at position 1, i.e., for 80.4\% cases, the model's best prediction was the corresponding incomplete triple's missing entity. We can interpret \textit{Hits@3} similarly. We translate the overall entity prediction result on CS40K to be sufficient such that the trained model can be improved and further used to infer unseen facts.

\begin{table}[!t]
%% increase table row spacing, adjust to taste
\setlength{\tabcolsep}{4.5pt}
\renewcommand{\arraystretch}{1.3}
\caption{Experimental evaluation of CS3K for entity prediction.}
\label{table:result:CS3K}
\centering
\begin{tabular}{c c c c c c}

Model &Hits@1$\uparrow$ &Hits@3$\uparrow$ &Hits@10$\uparrow$ & MR$\downarrow$ & MRR$\uparrow$ \\
\hline
TransH  & 26.8 & 50.5 & 65.2 & 32.34 & 0.414 \\
TuckER  & 64.3 & 70.5 & 81.21& 7.6 & 0.697 \\
\hline
\end{tabular}
\end{table}

\begin{table}[!t]
%% increase table row spacing, adjust to taste
\setlength{\tabcolsep}{4.5pt}
\renewcommand{\arraystretch}{1.3}
\caption{Experimental evaluation of CS40K for entity prediction.}
\label{table:result:CS40K}
\centering
\begin{tabular}{c c c c c c}

Model &Hits@1$\uparrow$ &Hits@3$\uparrow$ &Hits@10$\uparrow$ & MR$\downarrow$ & MRR$\uparrow$ \\
\hline
%TransH  & - & - & - & - & - \\
TuckER  & 73.9 & 75.9 & 80.4 & 202 & 0.75\\
\hline
\end{tabular}
\end{table}
\subsection{Use-case Description}\label{useCasesResults}
We query CTI-KG by forming an incomplete test triple as this is the formal approach to inferring information from a knowledge graph. The results are a set of predictions where each prediction has a confidence score associated with it. The confidence score of a predicted entity $e_i$ denotes the probability of $e_i$ being the accurate candidate for the incomplete triple. The predictions are sorted in descending order by this score, and we observe the top 10 predictions as potential predictions (\textit{Hits@10}).
\newline
\textbf{Case Study 1, Predicting a malware family:} A security analyst wants to identify which malware family is associated with the indicator $intel-update[.]com$ in a CTI-KG. The following query is posed to the CTI-KG as an incomplete triple: $\langle intel-update[.]com, indicates, ? \rangle$. Here, the accurate malware family associated with the domain $intel-update[.]com$ is named Stealer. We refer the reader to the FireEye report titled `Operation Saffron Rose'\cite{operationSaffronRose} for validating this fact. This CTI was not included in the training corpus. However, the training set had some triples involving Stealer, such as: $\langle office.windowsessentials[.]tk, indicates, Stealer\rangle$ and $\langle Saffron\_Rose,	involvesMalware,	Stealer\rangle$.
\newline
The model learns latent features of Stealer, its indicators of compromise, and other related entities (e.g., campaigns that involved Stealer) present in the training set. Table \ref{table:result-predictions} shows the ordered set of predictions returned by the model, and the true answer is ranked \#1. This is a significant result as it can simplify the job of an analyst to a great extent by narrowing down the scope for further analysis. Starting with this set of 10 results with different confidence scores, the analyst finds the accurate information in the second item. By definition, the \textit{Hits@3} and \textit{Hits@10} metrics increase as the true entity exists in the top 3 and certainly in the top 10. As seen in table \ref{table:result:CS40K}, we can say 75.9\% times the true entity will appear in the top 3 predictions of the model.
     \newline
\textbf{Case study 2, Inferring Attack Target:} Note that CTI-KG is built by extracting semantically rich information on a variety of attacks. We, therefore, constructed CTI-KG exclusively for android malware. We expected better inferences as the corpus was specific to one category of attacks. A security analyst queries CTI-KG to learn about the region impacted by a type of attack. The following query is posed to the CTI-KG in the form of an incomplete triple: $\langle $\textit{``coronavirus-themed attacks''}$, targets, ?\rangle$. The training set had some triples on  \textit{``coronavirus-themed attacks''}. Note that the training set contains facts from different CTI reports, and the terms \textit{``coronavirus-themed attacks''} and `targets' refer to the same attack category. The model learns latent features of these entities and relationships during training and later uses these features to infer a set of regions highly probable to involve the attack. Table \ref{table:result-predictions} demonstrates the inferences made by this query. We can see that the true answer (India) is ranked \#1. This result contributes to increasing the \textit{Hits@10} metric (since the rank is within the top 10), as well as to \textit{Hits@1} and \textit{Hits@3}. We obtain an intuition about what the Hits@\textit{n} metric on the overall test dataset represents. \textit{Hits@10} score reported in table \ref{table:result:CS40K} can be interpreted that 80.4\% time the true entity would appear in the top 10 inferences. We elaborate on this analysis in the following section.

\begin{table*}[t]
\centering
\caption{Detailed result of Inferring Entities. Malware hashes in the table: *01da7213940a74c292d09ebe17f1bd01,**01da7213940a74c292d09ebe17f1bd01}
\label{table:result-predictions}
\begin{tabular}{c| c c  | c c }
\hline
 &\multicolumn{2}{c|}{Case study 1: Inferring a Malware Family}& \multicolumn{2}{c}{Case study 2: Inferring Attack Target}\\
  & \multicolumn{2}{c|}{$\langle intel-update[.]com, indicates, ? \rangle$}&\multicolumn{2}{c}{$\langle $\textit{``coronavirus-themed attacks''}$, targets, ?\rangle$}\\
 \hline
Rank & Inferred Entity & Confidence score& Inferred entity& Confidence score\\
\hline
1  &  \textbf{Stealer}  &  \textbf{0.5769}  &  \textbf{India}  & \textbf{0.8683} \\
  2  & 	A malware hash* &  	0.5694  & recorded future & 0.8634  \\
  3  & A malware hash**  &  0.5679  &  issuemakerslab & 0.2972 \\
  4  & Gholee  &  0.5679 &  google  & 0.1873  \\
  5  & 200.63.46.33  &  0.5668 & China  & 0.1662  \\
 \hline
 \multicolumn{5}{l}{\footnotesize Bold text denotes the true entity for the corresponding test triple}\\
\hline
\end{tabular}
%\afterpage{}

\end{table*}

\begin{table*}[t]
\centering
\caption{Inferring Entity for different Data sets}
\label{table:result-benchmark3}
\begin{tabular}{c| c c  | c c | c c | c c }
\hline
 &\multicolumn{2}{c}{FB15K}& \multicolumn{2}{c}{WN18}
 & \multicolumn{2}{c}{FB15K-237}& \multicolumn{2}{c}{WN18-RR}\\
 \hline
Model & Hits@10$\uparrow$& MRR$\uparrow$ & Hits@10$\uparrow$ & MRR$\uparrow$& Hits@10$\uparrow$& MRR$\uparrow$ & Hits@10$\uparrow$ & MRR$\uparrow$\\
\hline
TransE & 44.3&0.227&75.4&0.395 &\textbf{32.2}&\textbf{0.169}&47.2&0.176\\
TransH&45.5&0.177&76.2&0.434&30.9&0.157&46.9&0.178\\
TransR&\textbf{48.8}&\textbf{0.236}&\textbf{77.8}&\textbf{0.441}&31.4&0.164&\textbf{48.1}&\textbf{0.184}\\
\hline
DistMult& 45.1 &0.206&80.9&0.531&30.3&0.151&46.2&0.264\\
ComplEx&43.8&0.205&\textbf{81.5}&\textbf{0.597}&29.9&0.158&46.7&\textbf{0.276}\\
TuckER&\textbf{51.3}&\textbf{0.260}&80.6&0.576&\textbf{35.4}&\textbf{0.197}&\textbf{46.8}&0.272\\
\hline
\multicolumn{9}{l}{\footnotesize The top three rows are models from the translation-based approach, rest are from tensor factorization-based approach}\\
\multicolumn{9}{l}{\footnotesize Bold numbers denote the best scores in each category}\\
\hline
\end{tabular}
\end{table*}

\section{Related Work}\label{relatedWork}
Cyber threat knowledge discovery and analysis allows security analysts to mine threat and machine-readable attack artifacts such as a list of indicators of compromise-- traffic signatures, malicious IP addresses, virus signatures, malicious URLs, and domains that organizations download and install into their firewall\cite{liao2016acing,husari2017ttpdrill,zhu2018chainsmith}. These artifacts are routinely used to prepare machine-learning models that perform predictive security analytics\cite{sun2018data,zhu2016featuresmith,husari2017ttpdrill}. Much of existing research may not continue to help prevent further attack events\cite{liao2016acing} as (a) attack characteristics change, \textit{e.g.}, malware samples frequently get repackaged\cite{caballero2011measuring}, (b) malicious domains do not remain active for a long time\cite{kuhrer2014paint}, or (c) the analysts, for example, may not connect the evidence they see in their internal network log alerts to an ongoing, more extensive, external attack campaign.
\par
Security analysts deal with hundreds of thousands of false alerts daily due to ML models not finding a pattern in their learning model. The latest work is gradually moving towards proactive prediction\cite{sun2018data} that captures the semantics behind audit data. Further advances have been made\cite{zhu2018chainsmith} to automatically analyze unstructured text to generate detectable patterns that consist of combinations of IoCs. However, there is no known significant effort research similar to our work; however, there is plentiful work on cyber threat knowledge for internal log networks\cite{zhu2018chainsmith,zhu2016featuresmith,satvat2021extractor}. Related work\cite{barnum2012standardizing} derives threat actions from reports and maps them to attack patterns with pre-defined ontology or machine learning models. While\cite{zhu2018chainsmith} work provides a good starting point towards extracting threat indicators, they miss out on the context in the form of the relationship between CTI concepts. Prior research \cite{rastogi2020malont,christian2021ontology} established that characterization of the threat from the attack and the defense perspective is crucial.

\section*{Acknowledgment}
This work was supported by IBM Research through the AI Horizons Network research grant.
\bibliographystyle{IEEEtran}
\bibliography{mybib}
% that's all folks
\end{document}